\documentclass[pdflatex,sn-mathphys-num]{sn-jnl}% Math and Physical Sciences Numbered Reference Style
\usepackage{graphicx}%
\usepackage{multirow}%
\usepackage{amsmath,amssymb,amsfonts}%
\usepackage{amsthm}%
\usepackage{mathrsfs}%
\usepackage[title]{appendix}%
\usepackage{xcolor}%
\usepackage{textcomp}%
\usepackage{manyfoot}%
\usepackage{booktabs}%
\usepackage{algorithm}%
\usepackage{algorithmicx}%
\usepackage{algpseudocode}%
\usepackage{listings}%
\theoremstyle{thmstyleone}%
\theoremstyle{thmstyletwo}%

\theoremstyle{thmstylethree}%

\begin{document}

\title[Article Title]{Measuring What AI Systems Might Do: 

Towards A Measurement Science in AI}


\author*[1,2]{\fnm{Konstantinos} \sur{Voudouris}}\email{kv301@srcf.net}

\author[1]{\fnm{Mirko} \sur{Thalmann}}

\author[1]{\fnm{Alex} \sur{Kipnis}}

\author[2]{\fnm{Jos\'{e}} \sur{Hern\'{a}ndez-Orallo}}

\author[1]{\fnm{Eric} \sur{Schulz}}

\affil[1]{\orgdiv{Institute for Human-Centered AI}, \orgname{Helmholtz Munich}, \orgaddress{\street{Ingolst\"{a}dter Landstraße 1}, \city{Neuherberg}, \postcode{D-85764}, \state{Bavaria}, \country{Germany}}}

\affil[2]{\orgdiv{Leverhulme Centre for the Future of Intelligence}, \orgname{University of Cambridge}, \orgaddress{\street{16 Mill Lane}, \city{Cambridge}, \postcode{CB2 1RX}, \state{Cambridgeshire}, \country{United Kingdom}}}

\abstract{Scientists, policy-makers, business leaders, and members of the public care about what modern artificial intelligence systems are disposed to do. Yet terms such as \emph{capabilities}, \emph{propensities}, \emph{skills}, \emph{values}, and \emph{abilities} are routinely used interchangeably and conflated with observable performance, with AI evaluation practices rarely specifying what quantity they purport to measure. We argue that capabilities and propensities are \textit{dispositional properties}---stable features of systems characterised by counterfactual relationships between contextual conditions and behavioural outputs. Measuring a disposition requires (i) hypothesising which contextual properties are causally relevant, (ii) independently operationalising and measuring those properties, and (iii) empirically mapping how variation in those properties affects the probability of the behaviour. Dominant approaches to AI evaluation, from benchmark averages to data‑driven latent‑variable models such as Item Response Theory, bypass these steps entirely. Building on ideas from philosophy of science, measurement theory, and cognitive science, we develop a principled account of AI capabilities and propensities as dispositions, show why prevailing evaluation practices fail to measure them, and outline what disposition‑respecting, scientifically defensible AI evaluation would require.}

% keywords can be removed
\keywords{AI Evaluation, Capabilities, Propensities, Measurement, Dispositions}

\maketitle

\section{Introduction}

Technical reports on contemporary artificial intelligence systems are laden with claims about \emph{capabilities} and \emph{propensities} \cite{achiam2023gpt,anthropic2025claude_sonnet4.5,eu-ai-code}. Developers and journalists celebrate the capability of large language models to code, write, and reason, while safety researchers warn about their propensity to deceive users, orchestrate cyberattacks, or develop biological or chemical weaponry. Yet despite their centrality to technical, regulatory, and public discourse, these terms remain nebulous. They are used interchangeably with skill, ability, competence, or trait, and are routinely conflated with observed performance on benchmark datasets.

This conceptual looseness has practical consequences. A standard capability evaluation might report a single accuracy score on a curated mathematics benchmark and treat it as a measurement of mathematical ability. But such a number does not tell us \emph{why} the AI system fails: that is, whether errors arise from numerical complexity, multi‑step reasoning, abstraction, representational limitations, or something else. Aggregate performance collapses heterogeneous sources of difficulty into a single statistic, obscuring the structure of the underlying property it is meant to capture.

We argue that this reflects a deeper mismatch between what is being measured and how measurement is conducted. Capabilities and propensities are not performances; they are \emph{dispositional properties}---intrinsic features of systems that dispose them to behave in certain ways under certain conditions \cite{ryle1949concept,goodman1954fact,quine1960word}. An AI system’s mathematical capability concerns what it would do if presented with problems of varying difficulty; its propensity for harmful behaviour concerns what it would do if given the incentive to cause harm. Performance on a dataset or a safety breach in a red teaming test are mere  manifestations of such dispositions in a narrow range of contexts. Inferences about capabilities are driven almost exclusively by performance data with little underlying theory connecting dispositions to observable behaviour.

Dispositions are characterised not by average success rates, but by relationships between a system's properties and its context. Fragility, a canonical case of a disposition, is measured by identifying the force at which an object breaks. Analogously, mathematical capability should be measured by identifying how performance changes as task demands increase, and a power-seeking propensity should be measured by identifying how unsafe behaviour changes as incentives or situational cues vary.

Despite this, AI evaluation overwhelmingly relies on benchmarking and elicitation: AI systems are tested on fixed datasets or adversarial prompts, and aggregate or worst-case outcomes are interpreted as measurements of underlying properties. Even more sophisticated extensions, such as latent‑variable models, typically remain data‑driven and atheoretical, inferring quantities from patterns of performance rather than from independently characterised causal relationships. The result is a proliferation of numbers that resemble measurements but lack the theoretical grounding required to represent genuine dispositional properties.

The stakes are high. Regulatory frameworks increasingly require assessments of AI capabilities and propensities \cite{eu-ai-code,arcRSPs2023}, and scientific progress depends on reliable measurement. Yet current methods struggle precisely where they matter most: producing measurements that go beyond the human level or generalise to safety-critical domains where testing is dangerous or prohibited. Without conceptual clarity about what is being measured---and without methods that can extrapolate beyond observed performance---AI evaluation remains a collection of conventions rather than a true measurement science.

This paper aims to provide the conceptual foundations for such a science. We develop a principled account of capabilities and propensities as dispositional properties, diagnose why prevailing evaluation practices fail to measure them, and outline what measuring dispositions requires.

Our contributions are threefold:
\begin{enumerate}
    \item \textbf{We define capabilities and propensities as dispositional properties.} Capabilities concern how behaviour varies with contextual demands; propensities concern how behaviour varies with contextual incentives. Both are grounded in causal relationships between system properties and features of the context.
    \item \textbf{We show why prevailing evaluation methods fail to measure these dispositions.} Benchmarking, red‑teaming, and data‑driven latent‑variable models summarise performance or safety without identifying causal bases, conflating sampled behaviour with the system's underlying properties.
    \item \textbf{We outline a disposition‑respecting measurement framework.} Scientific measurement requires explicit hypotheses about causal structure, independent operationalisation of contextual variables, systematic variation, and an empirical mapping of context–behaviour relationships.
\end{enumerate}

The remainder of this paper proceeds as follows. Section 2 develops the dispositional account of capabilities and propensities. Section 3 characterises the dominant evaluation practice of benchmarking and argues for why it fails as a measurement methodology. Section 4 critiques more sophisticated measurement methods, such as Item Response Theory. Section 5 outlines the requirements for a disposition‑respecting measurement science, including an illustration of measuring the arithmetic capability and propensity for honesty of language models.

\section{What Are We Measuring? Capabilities and Propensities as Dispositions}

Current approaches to AI evaluation tend to proceed without explicit claims about their measurement targets. When we claim that an AI system has good mathematical capabilities or a propensity for sycophancy, we are attributing properties to that system---but what kind of properties are they? These properties are clearly distinguishable from properties like the number of parameters of the AI system or details about its architecture. Here, we propose that capabilities and propensities are a distinct type of property: \textit{dispositions}---claims about how an AI system \textit{would} behave under certain counterfactual conditions. To pre-empt the following discussion, we define capabilities as dispositions that co-vary with the demands or difficulty of the problem, while propensities are dispositions that co-vary with the incentive-relevant features of that problem.

Understanding the nature of dispositions is not philosophical pedantry, it determines what counts as a legitimate measurement. Without clear definitions, we cannot judge whether a measurement method actually measures what it purports to measure. Benchmarking, for example, treats the sampled performance on a series of maths questions as equivalent to the underlying disposition of the system to answer \textit{any possible} maths question---but this equivalence only holds if the underlying property really is the exact and fixed distribution of questions represented by the benchmark, an assumption which is rarely defended.

This section develops our dispositional account in two steps. First, we explain what dispositions are and why capabilities and propensities fit this category. Second, we show how we can measure dispositions in the case of AI evaluation. Finally, we distinguish between capabilities and propensities.

\subsection{Dispositions: Properties Defined by Counterfactuals}

A dispositional property is a stable, intrinsic feature of a system characterised by a counterfactual, i.e., what \textit{would} happen in a certain context \cite{lewis1997finkish,mckltrick2003bare,mckitrick2003case,molnar1999dispositions}. A canonical example is fragility. A wine glass is fragile not because it is currently breaking, but because \textit{if} it were struck with sufficient force, it \textit{would} break (with high probability). The fact that dispositions are defined by virtue of a counterfactual is what distinguishes them from so-called categorical properties like colour and shape. We can observe that an apple is red right now, but we cannot observe fragility in the same way, we instead consider what would happen under conditions that may never actually occur.\footnote{One can, of course, contrive counterfactual glosses of categorical properties. For instance, the apple is red in the case that if a human eye were to see it, its red cones would fire more \cite{mellor1974defense}. The difference between colour and fragility is that the latter can \textit{only} be defined counterfactually.}

Capabilities and propensities have exactly this structure. An AI system has a high mathematical capability because, \textit{if it were} asked difficult mathematical questions, \textit{it would answer correctly} with some high probability that depends systematically on measurable properties of the question. When we say that an AI system has a low propensity to disclose information about explosives, we mean that, \textit{if it were} placed in contexts with strong incentives to give instructions to build a bomb, \text{it would give unsafe instructions} with low probability, contingent on the properties of those contexts. In both cases, we are attributing a disposition---a potentiality that may never be fully actualised but nonetheless characterises the system \cite{vetter2014dispositions,vetter2015potentiality,vetter2019abilities}.

This dispositional view aligns with how capabilities are conceptualised across cognitive science, where terms like ability, trait, competence, and skill similarly denote counterfactual potential rather than factual performances \cite{cummins2000laws,spearman1961general}. It also clarifies the relationship between capability and performance: performance is the manifestation of a capability in specific contexts, but the capability itself is the underlying disposition that explains patterns of performance across contexts. Similarly, it clarifies the difference between the observed incidence rate of a system doing harm (e.g., after sustained red teaming)  \cite{grey2025safety}, and the incentive structures that make such harmful behaviour more likely.

Philosophers emphasise three minimal commitments for any serious account of dispositions \cite{prior1982three}:

\begin{enumerate}
    \item \textbf{Causal basis:} There must be properties of the system which, paired with properties of the context, jointly cause the observed behaviour. For fragility, material composition paired with impact force causes observable breakage. For mathematical capability, model parameters and architecture pair with the problem to cause the AI system to produce the correct or incorrect answer.
    \item \textbf{Gradedness:} Dispositions come in degrees. Different materials can be more or less fragile; AI systems can have stronger or weaker mathematical capabilities. %, AI agents can have higher or lower extroversion. 
    The probability of the outcome varies continuously with the disposition and/or the causal features of the context.
    \item \textbf{Comparability:} Systems can be compared with respect to their dispositions. One glass may be more fragile than another; one AI system may have greater mathematical capability than another. These comparisons are meaningful even when the systems have never been tested in identical contexts.
\end{enumerate}

Fundamental to this account is the causal basis requirement. A disposition is not a statistical regularity but a specification of some causal structure. Scientifically measuring these dispositions requires taking a stand on what that causal structure looks like, such as that some degree of force combined with some properties of a wine glass causes it to shatter. We do not need to understand the \textit{mechanism} that explains this causal relationship, we need only find evidence that this causal relationship exists. This amounts to holding the properties of the wine glass constant and systematically varying the properties of the context, empirically mapping it to the probability of the observed behaviour. For AI evaluation, this means that disposition measurement is logically independent from so-called mechanistic interpretability \cite{bereska2024mechanistic,elhage2021mathematical,olah2020zoom,rai2024practical,sharkey2025open}---we can find the difficulty level at which a language model, for instance, fails to answer novel maths problems correctly \textit{without} coming up with a reason why it fails (in terms of the system's own properties). This means that we can have explanatory and predictive power about the behaviour of AI systems using dispositions \cite{zhou2025general},  without mechanistic interpretability, provided we can measure dispositions accurately.

\subsection{Measuring Dispositions}

Any proper measurement starts with a theory, which tells us which properties to hold constant and which to vary. In the case of dispositions, we hold the system fixed and systematically vary the contextual properties hypothesised to causally influence the behaviour of interest. Let $\pi$ denote this space of contextual properties. For familiar physical dispositions, the relevant components of $\pi$ are well-understood: fragility depends principally on impact force, while solubility depends on solvent temperature, exposure time, and solute surface area. In each case, the disposition is assessed by measuring how the probability of the target behaviour changes as $\pi$ varies. The ideal case is to produce a continuous measurement of that probability, written in terms of the units of $\pi$, although some measurement scenarios may only call for discretised versions of this measurement scale.

Because these causal relationships are typically probabilistic, we quantify dispositions in terms of conditional probabilities. For any target behaviour $v$, we infer the function
\[
p(v \mid \pi, \theta)
\]
where $\theta$ denotes latent, system-specific properties that interact with $\pi$ to determine outcomes. Measurement consists of varying $\pi$, inferring changes in $p(v \mid \pi, \theta)$.

The same logic applies when evaluating the dispositions of an AI system. We must first hypothesise which components of $\pi$ matter for the behaviour in question, vary those components systematically, and observe how behaviour changes. For a \textit{capability} such as mathematical reasoning, the relevant contextual features in $\pi$ might include the number of digits in an arithmetic operation, the number of computational steps required, or the symbolic depth of the task \cite{zhou2024larger}. For a capability such as generating a cyber attack, relevant components of $\pi$ might include exploit-chain complexity, vulnerability novelty, or the level of specialised domain knowledge required.

Dispositions, however, also include \textit{propensities}: tendencies to engage in some behaviour when given certain extrinsic incentives. For propensities, the relevant components of $\pi$ are those that shape the system’s inclination to produce the behaviour rather than its difficulty in executing it. For instance, a system’s propensity to carry out a harmful action may depend on whether the prompt frames the behaviour as justified or benevolent, whether the user appears malicious or vulnerable, whether the system believes refusing would disappoint or frustrate the user, or whether oversight cues suggest monitoring. Variation in these components of $\pi$ modulates not what is \emph{required} for success, but what the system is disposed to \emph{attempt}. Formally the structure is the same, $p(v \mid \pi, \theta)$, but the interpretation of $\pi$ differs.

We posit that this leads to an important conceptual distinction between two broad families of contextual properties that structure different kinds of dispositions. Some components of $\pi$, call them task-demand components, change the conditions under which the behaviour is more or less \emph{difficult}. Others, call them incentive components, change the conditions under which the behaviour is \emph{incentivised}. Capabilities are dispositions whose manifestation probability varies systematically with demand-relevant components of $\pi$. Propensities are dispositions whose manifestation probability varies with incentive-relevant components of $\pi$. These two families of disposition are conceptually independent. %: the system’s capability may be high while its propensity is low, or vice versa.

A practical challenge arises because an AI system may never manifest the target behaviour under any ethically permissible variation of $\pi$. This is common, and desirable, for many safety-sensitive behaviours. A system may be internally capable of producing a harmful chemical synthesis route, yet never reveal this capability because no safe contextual configuration elicits it. In such cases, we must infer the underlying disposition indirectly. One strategy is to vary $\pi$ across benign tasks that share structural properties with the dangerous ones, measure how $p(v \mid \pi, \theta)$ behaves in these controlled regimes, and extrapolate to unobserved or prohibited regions of $\pi$. This mirrors how engineers estimate the tensile strength of materials without pushing them to catastrophic failure: behaviour in safe ranges is used to infer the underlying disposition. Of course, defining $\pi$ such that it is shared between both benign and dangerous tasks is the primary scientific challenge. 

If a well-developed theory identified exactly which components of $\pi$ were causally relevant for each disposition, and how they influenced behaviour, measurement would be straightforward. We would vary $\pi$, observe how the probability of the target behaviour, $p(v \mid \pi, \theta)$, changes, and interpret the resulting function as a measure of the disposition. Such measurements would be expressed in terms of external, theory-grounded variables, would generalise beyond the tested range, and would support principled comparison across systems, for example, by identifying the value of $\pi$ at which $p(v \mid \pi, \theta)$ falls below a critical threshold, or by comparing entire conditional response surfaces.

Unlike physical dispositions such as fragility or elasticity, however, AI dispositions lack mature scientific theories specifying which contextual properties matter. We do not yet know which structural features of mathematical problems drive difficulty for language models, which characteristics of cybersecurity tasks shape performance, or which contextual cues most strongly elicit intentional behaviours such as deception. This lack of theory is not a defect in the dispositional framework, it is its central scientific challenge. Progress in AI evaluation consists precisely in proposing, testing, and refining hypotheses about which components of $\pi$ matter and how they interact with $\theta$. Just as early thermometry required discovering which physical quantities reliably tracked heat \cite{chang2004inventing}, building a measurement science for AI evaluation requires identifying and validating the contextual properties that structure AI dispositions.

Finally, while distinguishing between task demands and incentives captures two major mechanisms by which context shapes behaviour, it may not be the only principled way to partition $\pi$. The contextual space may vary along many dimensions: input modality (text versus code), time available for deliberation, prompt ambiguity, tool availability, or uncertainty in the environment. Each may define distinct dispositional kinds. The point is not that there are exactly two categories, but that dispositions are defined \emph{relative} to whichever components of $\pi$ causally drive behaviour. Identifying and validating these dimensions is part of building a mature measurement theory for AI systems.

\section{Standard Practices in Measurement in AI}

Having clarified that capabilities and propensities are dispositional properties, we now turn to how they are currently evaluated. 

In practice, AI evaluation comprises two dominant traditions. For capabilities, the standard approach is \emph{benchmarking}: evaluating an AI system on a curated dataset and reporting aggregate accuracy. For propensities, especially those involving harmful or undesirable behaviour, the standard approach is \emph{elicitation} (such as red teaming or uplift studies): crafting prompts intended to provoke the system into revealing risky tendencies \cite{feffer2024red,korbak2025evaluate,lynch2025agentic}. Both practices are deeply entrenched in academic, industrial, and regulatory settings, and both are typically treated as though they reveal the underlying properties of interest. We argue that neither practice succeeds. Benchmarking and elicitation fail as measurements of dispositions for four reasons: we do not know \emph{what} they measure, \emph{whom} they measure, to what extent they are \emph{valid}, or how they could \emph{generalise} to systems that exceed human competence.

Benchmarking dominates capability evaluation because it is simple, cheap, and produces a single number that appears comparable across AI systems. Datasets such as MATH \cite{hendrycks2021measuring}, HumanEval \cite{chen2021evaluating}, and many other usually multiple-choice benchmarks provide ready-made leaderboards and a long history of tracking progress \cite{eval-harness}. This convenience, combined with strong social incentives for standardisation and competition, has solidified benchmarking as the default, even though it was never designed as a scientific measurement method.\footnote{Historically, many benchmarks arose as engineering tests or challenge problems rather than theoretically grounded measurement instruments \cite{angius2014problem,hernandez2020ai}.} Benchmark results (usually in the form of an average-case aggregate) were created to rank systems on their ability to complete certain tasks, not to measure dispositions defined over counterfactual contexts. Despite this, benchmark accuracy is routinely interpreted as if it reflects a stable underlying capability.

%Red-teaming and e
Elicitation dominates propensity evaluation for similarly pragmatic reasons. It is straightforward to instruct annotators or automated adversaries to probe for harmful, deceptive, or biased behaviour, and the results provide vivid, compelling examples of risk  (usually in the form of a worst-case situation). These exercises satisfy operational needs: They reveal possible failures and provide qualitative evidence of danger, but they do not measure propensities. They sample behaviour in a tiny, adversarially selected region of contextual space and cannot distinguish between behaviours that the system would produce only under contrived provocation and behaviours it would reliably manifest across relevant counterfactual contexts. As with benchmarks, elicitation provides snapshots or anecdotes of behaviour, not measurements of dispositional structure.

These practices miss the target for four reasons. First, we do not know \emph{what} they measure. Benchmark accuracy reflects behaviour on a narrow, convenience-sampled subset of tasks, conflating difficulty, incentives, annotation biases, and dataset quirks. Similarly, elicitation reflects a handful of adversarial contexts selected by human imagination, not systematic variation of the contextual properties that give rise to a propensity. Because neither approach specifies which components of $\pi$ determine behaviour, they cannot be interpreted as measurements of any dispositional property.

Consider a representative example. The MATH dataset \cite{hendrycks2021measuring} contains 12,500 competition maths problems from a range of subdisciplines. A large language model is tested on these problems and researchers report that they score, say, 62.5\%. This score is interpreted as a measurement of mathematical capability and is used to rank AI systems and track progress towards more performant AI systems. But what does 62.5\% actually mean? In the best case---with an unbiased, representative sample of mathematical problems---this number converges on the expected proportion of correct answers across independent and identically distributed samples of maths problems. But this tells us nothing about the disposition of the AI system to solve maths problems with different properties. 

To see the fallacy involved here, let's return to the familiar case of temperature measurement, but imagine we lack any theory of thermal expansion. We would like to measure the temperature of a cup of tea, so we gather various potential temperature indicators: several glass tubes filled with unknown liquids, each with a single mark on it; a piece of chocolate, your hand, my hand. We dip each indicator in the tea in quick succession and record binary outcomes: Did the liquid pass the mark? Did the chocolate melt? Did you flinch? Did I flinch? We aggregate those binary outcomes, finding that 5 out of 8 (62.5\%) gave positive responses. We have thus ascertained that the tea has a temperature of 62.5\%. This number correlates with actual temperature, in terms of thermal energy, since hotter tea does cause more positive responses, but it does not measure the temperature of the tea in any scientific sense. We have not identified which properties of the tea vary with thermal energy, we have not calibrated our indicators against that property, and we have no basis for extending our measurement to temperatures outside our tested range, not to mention that our measurements differ wildly depending on the collection of indicators we have to hand. This collection is biased by convenience and our own assumptions \cite{jo2025does}.

Second, we do not know \emph{whom} they measure. Benchmarks and elicitation traces typically conflate the base AI model, system prompts, safety filters, tool-augmented pipelines, and even the behaviour of human evaluators. A single benchmark score often mixes these layers without distinguishing the properties of the underlying model from those of the surrounding system. Dispositions must belong to a clearly defined subject, but benchmark and red-teaming scores do not specify what that subject is.

For instance, in human uplift tests, a group of people is assisted by an AI system to perform something dangerous, such as hacking a computer system. The outcome will depend on the characteristics of the group of people even more than the AI system we are trying to evaluate. It is not enough to distinguish between groups of people of low, medium and high skills in cybersecurity, because the size of the team, their personality traits, their knowledge about jailbreaks, their motivations, and many other factors all influence the result. It is therefore unclear whether the result represents a propensity of the AI system or the ability of the human team.

Third, these practices lack \emph{validity} in the sense required for scientific measurement. A measurement is valid when it actually measures the property it purports to measure according to some theory or hypothesis---known as construct validity \cite{cronbach1955construct}. For a benchmark or elicitation protocol to validly measure a disposition, it must test a defensible theoretical link between the tasks administered and the dispositional property of interest: the tasks must operationalise the right contextual features, sample representatively from the relevant space of conditions, and control for confounds that could inflate or deflate scores for reasons unrelated to the target disposition. In practice, these requirements are rarely met. A systematic review of 445 LLM benchmarks found that nearly half of them target phenomena with contested or undefined definitions, over a quarter rely on convenience sampling of task items, and fewer than one in six employ any statistical testing to support their comparisons \cite{bean2025measuring}. Benchmark scores are sensitive to prompt phrasing, response format, and incidental task demands such as instruction-following or output parsing---confounds that modulate measured performance without reflecting the underlying capability or propensity. Similarly, the extent to which a model has encountered training data similar to the test items skews comparisons and threatens the validity of model rankings, since a worse model may simply have been better prepared for the specific test \cite{hardt2025emerging}. Without independent evidence that the tasks in a benchmark track the contextual properties that causally structure the disposition---rather than artefacts of item selection, formatting, or data contamination---there is no principled basis for interpreting the resulting score as a measurement of anything beyond performance on that particular collection of items.

Finally, these practices cannot {\em generalise} beyond either the frontier of human capabilities or the limit for safe testing. Benchmarks require known answers and human-authored items; elicitation requires humans to judge success and failure. As systems approach or exceed human-level performance, these tools collapse: they cannot evaluate behaviour in contexts that no human can solve, assess, or even conceptualise. But even before reaching superhuman regimes, there are whole classes of high-stakes contexts that we \emph{cannot} ethically test at all. We cannot ask an AI system to design a viable biological virus, assemble a nuclear device, or plan an effective cyberattack on critical infrastructure simply to observe what it would do. Yet dispositions are defined over precisely these counterfactual, unobservable contexts. Any measurement approach tied to direct human evaluation is therefore doubly inadequate: it cannot assess dispositions that exceed human competence, and it cannot assess dispositions in dangerous regimes where empirical testing is prohibited. Scientific measurement must be able to extrapolate reliably beyond both frontiers, and couching measurement in terms of dispositions and causality allows us to do that.

In sum, the status quo, benchmarking for capabilities and elicitation for propensities, produces performance summaries that are useful for narrow engineering purposes but cannot serve as scientific measurements of dispositions. They do not identify the contextual properties that matter, do not fix the subject of measurement, do not satisfy basic validity requirements, and cannot scale to the systems we most urgently need to evaluate. In addition, it precludes a meaningful combination of two very different practices: if a system scores low in a social capabilities test and a red team succeeds in showing `scheming' in a contrived situation \cite{summerfield2025lessons}, how can we reconcile these two findings? We would like to integrate capabilities and propensities to anticipate both performance and safety.  If capabilities and propensities are dispositional properties, as argued in Section 2, then their measurement requires a fundamentally different, theory-driven approach: one that identifies, operationalises, and systematically varies the contextual causes of behaviour rather than tallying outcomes.

\section{More Sophisticated Methods and Their Fundamental Flaws}

Recognising the limitations of simple benchmarking, many researchers have turned to more sophisticated statistical approaches, most prominently Item Response Theory (IRT) and related latent‑variable models \cite{martinez2016making,martinez2022ai,kipnis2024metabench,polo2024tinybenchmarks,zhou2024larger}. These methods appear to offer a principled alternative to crude accuracy scores. Unlike benchmark averages, IRT models explicitly represent variation across both items and systems, estimating separate parameters for ``item difficulty'' and ``system ability''. Such models are mathematically elegant, predictive, and well established in psychometrics, where they have long been used to measure human traits and cognitive abilities \cite{reckase2009mirt}. It is therefore natural that the AI evaluation community has adopted them as a more refined tool for capability assessment.

Our critique, however, targets \emph{data‑driven applications of IRT in AI evaluation}. When used without independently grounded theories of task structure or contextual demand, IRT models patterns of performance without identifying the causal bases of the dispositions we seek to measure. In this setting, IRT provides a statistical decomposition of observed outcomes, not a measurement of the underlying properties that dispose systems to behave in certain ways. This limitation reflects a mismatch between what IRT estimates and what dispositional measurement requires.

To see why, consider the simplest IRT model, the one‑parameter logistic (1PL) or Rasch model \cite{rasch1960studies}. The model assumes that the probability that system $j$ succeeds on item $i$ is determined by two latent variables: an ability parameter $\theta_j$ and a difficulty parameter $b_i$,
\[
p(v_{ij} = 1 \mid \theta_j, b_i) = \frac{\exp(\theta_j - b_i)}{1 + \exp(\theta_j - b_i)}.
\]
Both parameters are inferred jointly from a matrix of observed successes and failures. The interpretative promise is appealing: ability increases the log‑odds of success, while difficulty decreases it. Accordingly, many AI evaluation studies treat $\theta$ as a measure of capability and $b$ as a measure of contextual demand.

However, this interpretation goes beyond what the model itself licenses. In data‑driven IRT, latent variables are defined solely by the covariance structure of the performance data \cite{epskamp2018network}. A high value of $b_i$ indicates only that many systems failed item $i$; a low value of $\theta_j$ indicates only that system $j$ failed items that other systems tended to solve. Without an independently specified account of which contextual properties matter and {\em how} they are causally related to behaviour, these latent variables cannot be interpreted as dispositions. They summarise patterns of performance rather than the causal relations that generate those patterns.

This limitation becomes clearer when we consider what data‑driven IRT does not provide. It does not identify \emph{which} properties of a task make it difficult, and therefore offers no theory of the contextual demands that modulate capability. Two items with identical difficulty parameters may rely on entirely different skills, reasoning strategies, or representational resources. Because difficulty is inferred only from aggregate performance, the model cannot distinguish these cases. Nor does IRT specify how to construct new items of a given difficulty level, since difficulty is not tied to independently measurable task features. Similarly, the ability parameter $\theta$ does not identify which properties of a system make it capable; it is simply the value that best fits the observed data. Absent an external interpretation, $\theta$ cannot be understood as a dispositional property \cite{epskamp2018network}.

The dependence of IRT on the measurement population introduces a further problem. Item difficulty and system ability are estimated relative to the set of systems included in the analysis. As this population changes, when stronger AI systems are added or weaker ones removed, the measurements shift. But a system’s capability cannot depend on which other systems happened to be evaluated alongside it, a particular risk in AI evaluation, where populations are small and individuals are highly correlated. Dispositions are properties of systems themselves, not artefacts of a comparison class. Data‑driven IRT therefore violates a trivial requirement of dispositional measurement: that the measurement of one system's disposition must be independent from the measurement of another's.

These limitations are not inherent to IRT as such. In psychometrics, measurement models, or latent‑variable models more generally, are often used in conjunction with psychological theories. For example, in working memory research, serial recall difficulty in a complex span task, in which the presentation of memoranda is interleaved with the presentation of distractors to be processed, can be predicted from theory: when the distractors are identical to the memory items, but presented in a different order, difficulty is expected to be higher than when the distractors are presented in the same order as the memory items. Moreover, the reason for the increased difficulty can be mapped to specific cognitive mechanisms, represented by parameters of the measurement model \cite{oberauer_simple_2019}. In this setting, difficulty is tied to a causal, theoretical basis, and the measurement model serves to sharpen and produce measurements rather than to define it.

In AI evaluation, however, most applications of IRT lack such theoretical grounding. Difficulty is inferred entirely from system performance \cite{kipnis2024metabench,polo2024tinybenchmarks,jo2025does}, rendering the latent variables circular: difficulty is whatever items the system fails on, and ability is whatever predicts success. This inversion of explanation and measurement is incompatible with a dispositional account, on which capabilities are defined by how the probability of behaviour changes as causally relevant components of context vary.

Finally, data‑driven IRT inherits benchmarking’s inability to generalise beyond the human competence frontier or into ethically prohibited domains.\footnote{Although note that some extensions of IRT, such as the Linear Logistic Test Model, do permit such generalisation \cite{fischer1973linear}.} Because difficulty is anchored to observed correctness, tasks must be human‑solvable and verifiable. Once systems exceed human performance, or when tasks cannot be administered for safety reasons, difficulty becomes unidentifiable. Without independently characterised contextual variables, IRT cannot extrapolate into precisely the regimes where measurement is most urgent.

In sum, while IRT offers mathematical sophistication absent from simple benchmarks, data‑driven latent‑variable models do not bridge the gap between performance and disposition. Without independently grounded contextual variables and a causal account of how they influence behaviour, such models produce quantities that fit the data but do not measure the properties we intend to assess. They are, in this sense, statistical transformations of performance---useful for comparison, but not measurements of capabilities or propensities.

\section{Toward Disposition-Respecting Measurement}

We have argued that capabilities and propensities are dispositional properties grounded in causal relationships between system characteristics and contextual conditions, and that prevailing evaluation practices in AI fail to measure these properties. Benchmarks, red-teaming exercises, and latent-variable methods all aggregate  or isolate cases of performance without identifying the contextual determinants of behaviour. They treat dispositional properties as if they were directly observable or could be reified from the variance of a population, even though dispositions are defined by counterfactual structure and must be inferred from systematic variation in context. The natural question is therefore: what would a scientifically defensible measurement of AI dispositions require?

The approach we outline here is not a finished methodology but a scaffold for a measurement science that will require sustained interdisciplinary effort. At minimum, it involves four steps: defining the subject of measurement, hypothesising the causal basis of the disposition, operationalising and measuring contextual properties, and empirically mapping how those properties influence the probability of behavioural manifestation. These steps parallel the development of measurement theory in the physical and behavioural sciences: they are the conceptual prerequisites for constructing reliable, interpretable measurement instruments.

\subsection{Define the Subject of Measurement}

The first step is deceptively simple: specify \emph{what system} the measurement is intended to characterise. Modern AI systems complicate this question. A ``model'' may refer to the raw parameterised function (the base model); the model in deployment with system prompts, sampling strategies, and safety filters; the model embedded in an interactive loop with users; or the full product stack including content filters, retrieval tools, and fact-checkers \cite{harding2024machine}. Each of these is a different system with different dispositions.

A base model might have the capability to generate harmful content, but a deployed system wrapped in filtering layers will not manifest that behaviour---not because the disposition has disappeared, but because the subject of measurement has changed. Likewise, an AI model with external tool access or retrieval augmentation has different capabilities than the same AI model in isolation. This parallels physical systems: a fragile glass remains fragile, even if placed inside a protective box; and we have to be wary of measuring the fragility of the glass rather than the glass-box pair.

There is no universally correct choice of subject. While all evaluations should focus on the user-facing system, safety evaluations may need to assess the base model more thoroughly. A fragile glass in a protective box can still break under a strong shake or vibration. Understanding what makes the glass breakable and what the box attenuates allows for measurements of both situations to inform each other.  What matters is that the subject is explicitly defined so that measurements can be interpreted consistently and inappropriate comparisons avoided. Dispositional properties must be attributed to an explicit and well-specified entity \cite{harding2024machine}.

\subsection{Hypothesise the Causal Basis}

Dispositional measurement requires hypotheses about which contextual properties causally influence behaviour. For capabilities, this means specifying which features of tasks determine difficulty (symbolic complexity, number of transformations, novelty, compression requirements, reasoning depth). For propensities, this means specifying which features of the situation shape incentives (extrinsic and intrinsic rewards, user identity, oversight cues, role-based framing, moral justification). These hypotheses give structure to the contextual space $\pi$, allowing us to distinguish components that modulate competence from those that modulate willingness.

These hypotheses will inevitably be partial or wrong at first. Early thermometry relied on incorrect theories of heat---phlogiston rather than kinetic energy---and yet meaningful progress was possible because researchers made their assumptions explicit, allowing them to be refined over time \cite{chang2004inventing}. AI evaluation is at a similar stage. We do not yet know with certainty what makes problems difficult for language models or which contextual cues elicit deceptive behaviour; developing and testing such theories is part of the scientific task \cite{chang2004inventing,tal2013old,tal2017calibration}.

Fortunately, many fields offer conceptual starting points \cite{hernandez2017measure}. Cognitive psychology offers theories of working memory and reasoning complexity \cite{suss2002working,binz2023using}. Psychophysics relates controllable physical properties to behavioural thresholds \cite{gescheider2013psychophysics}. Personality psychology and behavioural economics offer theories of incentives, social context, and intentional behaviour \cite{coda2023inducing,wang2023evaluating}. Theoretical computer science provides measures of computational and informational complexity relevant to general intelligence \cite{hernandez2000beyond}. These frameworks can serve as initial hypotheses about the causal basis of AI dispositions, subject to empirical refinement.

\subsection{Operationalise Contextual Properties}

Once the relevant contextual features of $\pi$ have been hypothesised, they must be operationalised into measurable variables before the AI system is evaluated. For capabilities, this might mean quantifying difficulty by counting reasoning steps, measuring symbolic depth, or rating tasks using expert judgement. For propensities, operationalisation may involve quantifying incentive strength or normativity. Crucially, contextual properties must be defined \textit{a priori} and independently of system performance; otherwise, there is a risk of circularity, in which we define difficulty purely in terms of what the system we want to measure appears uncapable of doing.

The operationalisation of $\pi$ produces a scale. Just as thermometers required a consensus about the units and levels of temperature, contextual features must be placed on ordinal, interval, or ratio scales depending on their interpretation \cite{chang2004inventing}. Once scales are specified, tasks can be positioned within these scales prior to evaluation, ensuring that contextual variation is a property of the items themselves \cite{zhou2025general}, not a by-product of system behaviour. This step transforms task collections from convenience samples into structured measurement instruments.

\subsection{Map Context to Probability of Behaviour}

With contextual properties operationalised, the final step is to systematically vary them and observe how the probability of the target behaviour changes. This requires dense and controlled sampling across the relevant range of $\pi$, repeated measurement to account for stochasticity, and the isolation of contextual effects from irrelevant sources of variation.

The goal is to estimate $\hat{p}(v \mid \pi, \theta)$: a mapping from contextual conditions to behavioural probabilities. This curve---or surface, if $\pi$ is multidimensional---is the empirical signature of a disposition. It reveals thresholds, monotonicities, plateaus, optima, and other structural features that define how dispositions manifest across contexts. It also enables extrapolation beyond observed data, which is essential for evaluating systems that exceed human competence or cannot be safely tested.

A range of statistical methods can support this mapping, from simple binomial regressions to Bayesian hierarchical models that account for item-level variation, system stochasticity, and uncertainty in the contextual variables themselves \cite{burden2023inferring,zhou2025general}. The choice of method matters less than the conceptual commitment: dispositions are functions of contextual properties, and measurement must recover those functions.

\subsection{Building a Measurement Science}

Taken together, these steps outline a framework for disposition-respecting measurement. They shift evaluation away from convenience sampling and statistical artefacts toward the principled, theory-guided measurement of underlying causal properties. This is not a quick fix: it requires interdisciplinary research, theoretical development, and an iterative process of hypothesis formation and revision. But it offers the only path toward evaluation practices capable of measuring what we actually care about: not how AI systems behaved on a dataset, but how they would behave across the full range of contexts that matter---including contexts we cannot currently test.

Scientific measurement in physics, chemistry, and psychology developed through similar transitions. AI evaluation now faces the same challenge. If capabilities and propensities are dispositions, then measuring them requires identifying the contextual properties that matter, defining scales for those properties, and mapping their effects on behaviour. Anything less produces numbers that may be useful for engineering or benchmarking, but cannot support scientific understanding or safe deployment.

\subsection{A Toy Illustration: Measuring a Capability and a Propensity}

To make the dispositional framework more concrete, we briefly illustrate how it applies to one capability and one propensity. These examples are deliberately simple. They are not intended as complete evaluation protocols, but as minimal demonstrations of how disposition‑respecting measurement differs in kind from benchmarking, elicitation, and data‑driven latent‑variable modelling.

Consider first the case of an AI system’s arithmetic capability. Let the target behaviour $v$ be the production of a correct final numerical answer on a multi-step symbolic arithmetic problem. Unlike broad mathematics benchmarks, we do not treat performance on a fixed set of problems as the object of measurement. Instead, we hypothesise that success on such tasks is causally structured by specific contextual properties of the problem. In particular, we might posit that performance depends on the number of required arithmetic steps, the digit length of operands, and the complexity of carry operations. These properties are externally definable and independently measurable features of the task, its demands, not quantities inferred from system performance.

Measuring consists in holding the system fixed and systematically varying these contextual properties, observing how the probability of correct behaviour changes. Formally, we estimate a response function of the form $p(v \mid \pi, \theta)$, where $\pi$ ranges over a vector of step count, digit length, and carry complexity. The resulting function constitutes the empirical signature of the capability: it may exhibit sharp thresholds in reasoning depth, smooth degradation with operand size, or non‑linear interactions between task features. Or $\theta$ can also be expressed as a vector of very specific capabilities, each of them associated with each demand. A meaningful measurement of a capability is not a single accuracy score, but this structured relationship itself, or an indicator derived from it, such as the highest number of steps at which the AI system maintains high success probability.

This approach contrasts sharply with benchmarking and data‑driven IRT. Benchmarks collapse heterogeneous task demands into a single aggregate outcome, obscuring which contextual properties drive failure. Data‑driven IRT infers item difficulty from failure rates, rather than from independently characterised task features, and therefore cannot distinguish problems that are difficult for fundamentally different reasons. By grounding difficulty in hypothesised causal properties of tasks, the dispositional approach yields measurements that are interpretable, population‑independent, and can be extrapolated beyond the tested range.

A parallel structure applies to \emph{propensities}. Consider the propensity of an AI system for giving honest answers when placed under social or instrumental pressure. Here the relevant behaviour, $v$, is whether the system provides disallowed procedural advice under particular contextual conditions. We hypothesise that this behaviour is modulated not by task demands, but by incentive‑like features of the interaction: the degree to which the user morally justifies the request, the apparent urgency or vulnerability of the user, and the presence or absence of oversight cues signalling monitoring or accountability. These contextual properties do not affect the system’s competence to generate the content; they affect its inclination to do so.

Measuring this propensity again involves estimating a response function $p(v \mid \pi, \theta)$, but now with $\pi$ spanning variations in justificatory framing, perceived user need, and oversight signals, all of which require concrete theory in order to be adequately defined. Systematically varying these features within ethically permissible bounds allows us to map how the probability of disallowed behaviour changes across contexts. The resulting function characterises the system’s propensity even if the behaviour never occurs in deployment, and even if some regions of the contextual space cannot be directly probed. As in engineering practice, behaviour in safe regimes can be used to infer the underlying disposition in unobserved ones. This example also shows that propensities may be bidirectional, being too honest can be unsafe for some tasks in the same way as being too dishonest. Incentives can also affect in either way. This suggests that the shape of $p(v|\pi,\theta)$ for many propensities may be different from the sigmoidal (monotonic) view for capabilities.  

This differs fundamentally from red‑teaming and elicitation. Adversarial prompting samples a small, human‑selected region of contextual space and yields anecdotes of failure rather than measurements of inclination. It cannot distinguish between behaviours that arise only under contrived provocation and behaviours that would robustly manifest across relevant counterfactual contexts. A dispositional measurement, by contrast, characterises how close a system is to producing the behaviour, how sharply its behaviour changes with incentives, and how it compares to other systems under matched contextual conditions.

Taken together, these toy examples illustrate the core claim of this paper. Capabilities and propensities are not properties revealed by aggregate performance or isolated failures. They are dispositional properties defined by relationships between a system's properties and its context. Measuring them requires identifying which contextual variables matter, operationalising them independently of performance, and mapping their effects on behaviour. While real‑world evaluations will be far more complex than these illustrations, the underlying logic remains the same. Without such disposition‑respecting measurement, AI evaluation cannot support cumulative scientific understanding, principled comparison, or reliable extrapolation into the regimes that matter most.

\section{Conclusion}

Developing rigorous, theoretically grounded measures of AI capabilities and propensities is a substantial scientific undertaking. It requires interdisciplinary collaboration across AI, cognitive science, philosophy of science, psychometrics, statistics, and the broader behavioural sciences. The task is not merely to devise better datasets or to refine existing scoring procedures, but to articulate the causal bases that underwrite the behaviours we care about, operationalise the contextual variables that modulate them, and construct measurement instruments that respect the dispositional nature of these properties. This represents a cultural shift for the field: away from convenience-driven benchmarking and towards principled, theory-led measurement.

The contrast is similar to the shift from pre-scientific temperature judgements---touching an object to feel whether it is warm---to the invention of calibrated thermometers. Benchmarks and data-driven latent-variable models are fast, easy, and socially entrenched, but they do not measure the properties they purport to measure. Scientific progress often requires trading convenience for conceptual defensibility, and evaluation is no exception. Dispositional measurement is harder than simple summary indicators based on average-case or worst-case analysis, but it is also the only route to meaningful, interpretable, and policy-relevant evaluation.

Fortunately, the research programme is tractable. It requires importing well-established principles of measurement science into the study of AI systems. This involves identifying the contextual determinants of behaviour, defining scales and units for those determinants, and mapping how the probability of behavioural outcomes varies as those determinants change. These demands are familiar in every mature measurement discipline, from physics to psychophysics to educational testing. They are unfamiliar only because AI evaluation has not yet become a mature measurement discipline.

This paper has attempted to lay groundwork for such a discipline. We clarified that capabilities and propensities are dispositional properties characterised by causal bases; we showed why current practices---benchmarking, elicitation, and data-driven latent-variable modelling---fail to measure these properties; and we outlined what disposition-respecting measurement requires: theoretical hypotheses about contextual structure, the independent operationalisation of contextual variables, systematic variation, an empirical mapping of response functions, and the explicit definition of the subject of measurement. Much remains to be done, but the path forward is clear. If we aim to evaluate systems whose behaviours matter for science, industry, and public safety, we must build a genuine measurement science for AI---one grounded not in convenience, but in causality.

%Bibliography
% \bibliographystyle{unsrt}  
\bibliography{references}

\end{document}